\documentclass[envcountsect,english,runningheads,a4paper]{llncs}[2018/03/10]

\usepackage{amsfonts}
\usepackage{amsmath}
\usepackage{float}
\usepackage[utf8]{inputenc}

\usepackage{graphicx}
\usepackage{multirow}
\usepackage{graphicx}
\usepackage{enumitem} 
\usepackage{url}
\usepackage{xcolor}
\usepackage{hyperref}

\usepackage[all]{hypcap}

\usepackage{xspace}
\usepackage{diagbox}
\usepackage{siunitx}
\usepackage{caption}
\newcolumntype{P}[1]{>{\centering\arraybackslash}p{#1}}
\newcommand{\vspaceshrinker}{\vspace{-2mm}} 
\newcommand{\ie}{\textit{i.e., }}  
\newcommand{\eg}{\textit{e.g., }}  
\newcommand{\cf}{\textit{cf. }} 
\def\1{\textit{(i)}}
\def\2{\textit{(ii)}}
\def\3{\textit{(iii)}}
\def\4{\textit{(iv)}}

\def\hpvnospace{Hybrid PV}
\def\hpv{Hybrid PV }

\newtheorem{thm}{Definition}[chapter]

\begin{document}
\title{Proverum: A Hybrid Public Verifiability \\ and Decentralized Identity Management}
\titlerunning{Hybrid Public Verifiability and Decentralized Identity Management}
\author{Christian Killer \and Lucas Thorbecke \and Bruno Rodrigues \and \\ Eder Scheid \and Muriel Franco \and Burkhard Stiller}
\authorrunning{C. Killer et al.}
\institute{Communication Systems Group CSG, Department of Informatics IfI, \\ Universit\"at Z\"urich UZH, Binzm{\"u}hlestrasse 14, CH---8050 Z{\"u}rich, Switzerland \\\email{\{killer,rodrigues,scheid,franco,stiller\}@ifi.uzh.ch}}

\maketitle
\begin{abstract}
Trust in electoral processes is fundamental for democracies. Further, the identity management of citizen data is crucial, because final tallies cannot be guaranteed without the assurance that every final vote was cast by an eligible voter. In order to establish a basis for a hybrid public verifiability of voting, this work \1 introduces Proverum, an approach combining a private environment based on private permissioned Distributed Ledgers with a public environment based on public Blockchains, \2 describes the application of the Proverum architecture to the Swiss Remote Postal Voting system, mitigating threats present in the current system, and \3 addresses successfully the decentralized identity management in a federalistic state.

\end{abstract}

\begin{keywords}
  Hybrid Public Verifiability, Decentralized Identity Management, Remote Postal Voting
\end{keywords}

\section{Introduction}
Trust in the integrity of voting processes or final tallies is crucial for democracies. One key property to achieve this is verifiability, a fundamental building block for transparency, and establish trust in voting processes \cite{volkamer11}. \textit{E.g., public verifiability} allows anyone to verify the accuracy of the voting process and the final tally independently. Verifiability for voting is achieved by deploying Public Bulletin Boards (PBB), which store ballots and verifiable logs. Since PBBs are append-only data structures and publicly shared, anyone can verify information on them. 

Often verifiability mechanisms are based on applied cryptography and protocols, which are not easily understood by the public. Therefore, \textit{administrative verifiability} plays a crucial role. From an election's administrative perspective, \textit{administrative verifiability} is usually accomplished in analog processes with paper audit trails enabling manual recounts. Such mechanisms rely on redundancy (\eg the four-eye principle) to mitigate collusion and minimize trust placed in single entities. Therefore, \cite{Benaloh2008} argues whether the combination of \textit{public} and \textit{administrative verifiability} can reach publicly verifiable voting. The integrity of an election can only be assured, when authorized voters can cast valid ballots. 

In the context of electronic voting, the property of \textit{eligibility verifiability} denotes that anyone can check that each ballot published on a PBB was cast by a registered, eligible voter and at most one vote is tallied per voter \cite{krs10}. \textit{Unforgeability} states that only eligible voters can construct authorized ballots \cite{smy18}. These properties achieve more robust privacy and verifiability mechanisms for existing voting processes. In order to achieve \textit{unforgeability} and \textit{eligibility verifiability} simultaneously, it is necessary to deploy a suitable Identity Management (IdM), which assures the integrity of the citizen and electorate data. Thus, achieving a notion of \textit{public verifiability} is impossible without a reliable IdM. 

From a governmental perspective, the integrity of citizen identity data is crucial for formal processes, not only voting. Keeping citizen registries updated and providing identity-dependent processes (\eg the creation of Electoral Registers (ER)) are fundamental in a federalistic system, where authorities (\eg Cantons in Switzerland) remain control over their data and infrastructure. Deploying a secure IdM in a federalistic state prefers a decentralized approach while enabling the secure management and a privacy-preserving exchange of identity data. 

Therefore, IdM of citizens is fundamental for any government, and electronic means provide more efficient solutions to manage citizen data. \textit{E.g.,} Estonia provides a physical smart card, which can be used by Estonian citizens to authenticate themselves online and sign legally-binding documents \cite{e-estonia}. The development of Electronic Identity (eID) solutions was launched or is being developed specifically for Switzerland \cite{swissID}. Even though these proprietary solutions are legally obliged to grant users control over their personal data, the centralized and opaque nature of deployments indicates otherwise. In turn, novel approaches toward a Self-Sovereign Identity (SSI) appeared with the promise of empowering users with control over their data. SSI solutions gained a suitable platform for efficient SSI solutions \cite{sovrin} due to public Blockchains (BC), which offer an alternative to remedy ownership and access control issues. Users can manage credentials stored on the BC, while also improving transparency, security, verifiability, and interoperability around the processes of IdM as a whole. Thus, simplified management of trusted identity information is reached, enabling government agencies to access, share, and use sensitive data, while maintaining integrity.

Thus, this work here proposes the Proverum architecture to address the lack of public verifiability in voting processes and introduces a new definition of Hybrid Public Verifiability (\hpvnospace). Proverum combines a public and private environment, allowing for tamper-proof audit trails, published by peers operated by authorized entities and independently verifiable by anyone. The private environment serves as a private permissioned distributed network formed by authorities. The public environment's deployment depends on the specific requirements of public verifiability being designed by either \1 publishing to a public BC, \2 deploying a public permissioned BC, or \3 granting partial read access to a subset of the private environment. Proverum defines a practically feasible approach offering \hpv for voting, allowing the public to verify data within a public environment, while maintaining a privacy-preserving, verifiable audit-trail within the private environment. While the Proverum architecture can be applied beyond voting, \eg for electoral processes or a sharing of information in the health-care sector (\eg reporting verifiable infection numbers of a pandemic disease outbreak), this paper's focus is on applying Proverum to the Swiss Remote Postal Voting (RPV) system. The current prototype focuses on the implementation of a private environment, modeling different Swiss municipalities. The source code is available in \cite{proverum}. 


The remainder of this paper is organized as follows. Relevant background and related work is presented in Section~\ref{sec:bgrw}. While Section~\ref{sec:proverum} provides all Proverum design aspects, Section~\ref{sec:implementation} details the prototype, following with Section~\ref{sec:usecases} on different use cases. Section \ref{sec:evaluation} renders a detailed evaluation, while finally Section~\ref{sec:conclusion} draws major conclusions.
\section{Background and Related Work}
\label{sec:bgrw}
Following the outline of relevant background on Blockchains (BC) and Distributed Ledgers (DL), the same is provided for Identity Management (IdM) and the overview on the Swiss Remote Postal Voting (RPV) approach.

\subsection{Blockchains and Distributed Ledgers}
\label{subsec:bgrw:bc}
\begin{figure}[!ht]
    \centering
    \includegraphics[width=0.9\textwidth]{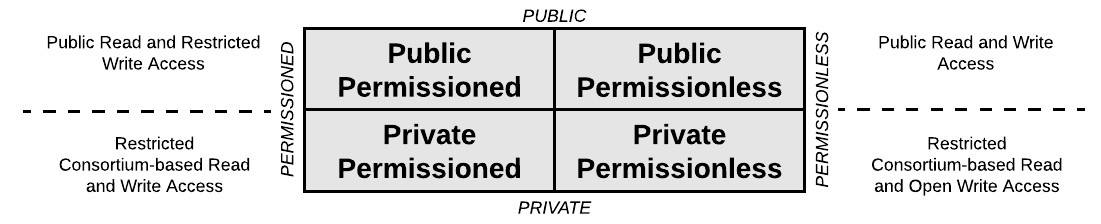}
    \caption{BC and DL Deployment Types, based on \cite{scheidBCcontributions}}
    \label{fig:bcdeployment}
    \vspaceshrinker
\end{figure}

A BC is an immutable backward linked-list, formed by blocks of transactions, which are maintained within a distributed network, governed by peers following a consensus mechanism. Four main development types of BCs are classified in Figure~\ref{fig:bcdeployment}. Each quadrant contains a deployment type: the \textit{x}-axis represents write permissions and the \textit{y}-axis read permissions. All permissioned BCs are better labelled DL, since they show one or more restrictive characteristics. 

\textit{Public Permissionless} BCs are the most prevalent deployment type, and most cryptocurrencies are implemented as such. Typically, these BCs serve as a tamper-proof and transparent platform for trustless exchanges, eliminating the need for a Trusted Third Party (TTP) as an intermediary \cite{bcInTheWings}. As outlined by Bitcoin, these deployments are open to \1 all participants regarding a read and a write access and \2 the participation in the consensus mechanism \cite{btc}. 

\textit{Public Permissioned} DLs offer a public read access, however, write permissions are restricted to a set of authorities. These DLs are suitable for a setting of multiple trusted authorities wanting to publish publicly verifiable data, accessible to anyone (\eg publishing hashes of, or encrypted votes in a Remote Electronic Voting (REV) system \cite{caiv}). 

\textit{Private Permissioned} DLs are often used in enterprise settings, where a consortium consisting of entities identified collaborate. Also, a \textit{private} DL can be used to establish transparency and verifiability among participants of a given process, while enabling the granular configuration of access control over data and allowing more performant consensus mechanisms to be deployed.

\textit{Permissioned} DL deployments basically introduce TTPs again. Prominent projects include Hyperledger Fabric (HLF) \cite{hlfdocs} or Corda \cite{corda-whitepaper}, which allow for a modular configuration of roles, access control, and data exchange channels.

While \cite{bcInTheWings} provides a BC and DL overview, specifically HLF and Corda as permissioned DLs are similar regarding the usage of certificate standards for IdM and different with respect to consensus mechanisms and underlying data structures. On one hand, HLF arranges all transactions in a DL. On the other hand, Corda sees all transactions as private. Each individual state is tracked under the supervision of a trusted, neutral third party notary, which orders and tracks these states to guarantee their validity and avoid double-spending. 

\subsection{Verifiability}
Typically, in paper-based voting systems (\eg the Swiss RPV systems), the voter cannot verify that her vote was correctly included in the final tally. She would have to trace the ballot through the process of placing it into the ballot box (or post box), emptying the box, anonymizing the ballots, and tallying the votes, which is unrealistic for a large number of voters. Therefore, verifiability in voting systems is not a binary concept \cite{Benaloh2008}. Verifiability notions depend on the sets of stakeholders, mechanisms, and assumptions in place, and therefore the following definitions serve as a reference for those contributions made in this work. For a systematic overview, please refer to \cite{cortier2016}.

\begin{thm}{Individual Verifiability (IV)}\\
A voting system has the IV property if a voter can verify that her ballot is in the recorded set of ballots and contains her intended vote \cite{jmp13}.
\end{thm}

\begin{thm}{Universal Verifiability (UV)}\\
A voting system has the UV property if anyone can verify that the voting result was tallied correctly, meaning that all valid ballots were included with their intended vote \cite{sk95}. 
\end{thm}

\begin{thm}{End-to-End Verifiability (E2E-V)}\\
    E2E-V is thmd similarly to IV and UV, but divided into three more specific properties. See \cite{brr15} for an in-depth definition.
        \begin{enumerate}
            \item \textit{Cast-as-Intended} verifiability requires that a voter can verify that her ballot contains the intended vote. For example, this is not obvious if votes are encrypted before sending them.
            \item \textit{Recorded-as-Cast} verifiability demands that a voter can verify that the voting system received and stored her ballot correctly. \textit{E.g.}, in case of a public list of votes, the voter can consult the list and check, if it contains her vote.
            \item \textit{Counted-as-Recorded} verifiability says that anyone can verify the correctness of the voting result, meaning that it includes all the recorded and valid ballots.
        \end{enumerate}
\end{thm}

E2E-V and the combination of IV and UV cover similar aspects. In research, usually either E2E-V or the combination of IV and UV is applied, but not both simultaneously. IV and Cast-as-Intended verifiability (CaIV) address the problem of a non-trusted voter platform in REV. CaIV needs to be addressed, if the voter's platform cannot be controlled and trusted, which is typically the case, if the voter is able to cast an electronic ballot remotely. IV and CaIV together assure that the voter can detect that her vote was compromised, most probably by malicious code on the voting device \cite{caiv}. While IV, UV, and E2E-V are of central importance for REV schemes, they are hard to prove in combination, but even harder to be deployed within an operational and practical RPV system. Therefore, the following notions of verifiability are necessary, when existing systems are evaluated, such as the Swiss RPV system.

\begin{thm}{Eligibility Verifiability}\\
A voting system has the eligibility verifiability property if anyone can verify that the voting result only contains votes from eligible voters, and only includes one vote per voter \cite{Kremer2010}.
\end{thm}

Eligibility verifiability is crucial for any voting system, independent of being based on paper ballots or operated electronically, since it assures the integrity of the final tally, making it verifiable that only eligible voters cast their vote once.

\begin{thm}{Administrative Verifiability (AV)}\\
\label{av}
A voting system offers AV, when election officials have means to protect against certain kinds of errors and fraud, typically accomplished with tools like paper audit trails that enable manual recounts and spot checks~\cite{Benaloh2008}.
\end{thm}

\begin{thm}{Public Verifiability (PV)}\\
\label{pv}
A voting system provides PV, when any individual can verify the accuracy of a tally regardless of any conspiracies of any size~\cite{Benaloh2008}.
\end{thm}

Based on these definitions, AV includes various processes implemented by trusted authorities, while PV requires cryptographic mechanisms to be deployed. In other words, AV determines real-world processes in operation and PV is based on an ideal, theoretical property and hard to deploy in practise.

\subsection{Identity Management}
\label{subsec:bgrw:idm}
Identity Management (IdM) determines a very active field of research and an economic sector. From a global perspective, a range of Single Sign-on (SSO) solutions and underlying standards exists. The most common protocols include OpenID and OpenID Connect (which extends OpenID with OAuth2 authorization) \cite{openID}, which define open and decentralized standards that let users choose from a variety of contributing identity providers to use an account elsewhere for authentication and authorization purposes. The emergence of BCs initiated alternative IdM approaches. For instance, Sovrin uses a public permissioned BC to form a distributed identity network \cite{sovrin}. Read access to Sovrin is public, while running a so-called \textit{validating node} within the network, which has to be authorized by the Sovrin Foundation. The initial code base of the Sovrin project originates from Hyperledger Indy, supervised by the Linux foundation \cite{hli}. The development of decentralized identity solutions addresses in new standards verifiable credentials (VC) \cite{vc} overseen by the World Wide Web Consortium (W3C) and the Decentralized Identity Foundation (DIF).

\subsection{The Swiss Remote Postal Voting (RPV) System}
\label{subsec:bgrw:pvpf}
The Swiss RPV is inherently built on External Service Providers (ESP) and trusted relationships among all parties involved. For eligible voters, the current process is hard to decipher and impossible to verify. \cite{pvpf} provides ($i$) an overview of the Swiss RPV system, ($ii$) a detailed insight into the process flow, and ($iii$) a respective risk assessment. This analysis and risk assessment in \cite{pvpf} provides critical Threat Events (TE) emerging during various stages of the voting process. The current RPV system offers benefits, too, such as its physical decentralization and the distribution of trust. Due to Switzerland's federal and decentralized structure, each Canton and municipality manages their respective jurisdictional electoral procedures autonomously. Cantons use centralized information systems to administer or transfer crucial data, \eg Electoral Registers (ER) or Web-based assistance tools, to transmit intermediate results \cite{sommer2019cyberrisks}. The deployment of a digitized REV potentially decreases the necessary amount of trust placed in institutions and people, shifting trust to transparent and verifiable processes instead~\cite{volkamer11}. For instance, various cryptographic mechanisms enable the electorate to verify different steps of these processes (\eg end-to-end verifiability, E2E-V).

\section{Proverum Architecture}
\label{sec:proverum}
As stated, the key goal of Proverum is to provide a feasible approach to provide \hpv for voting and electoral processes, which allows the public to verify data in a public environment, while maintaining a privacy-preserving and verifiable audit-trail in the private environment. 

\begin{figure}[!h]
    \centering
    \includegraphics[width=0.8
    \textwidth]{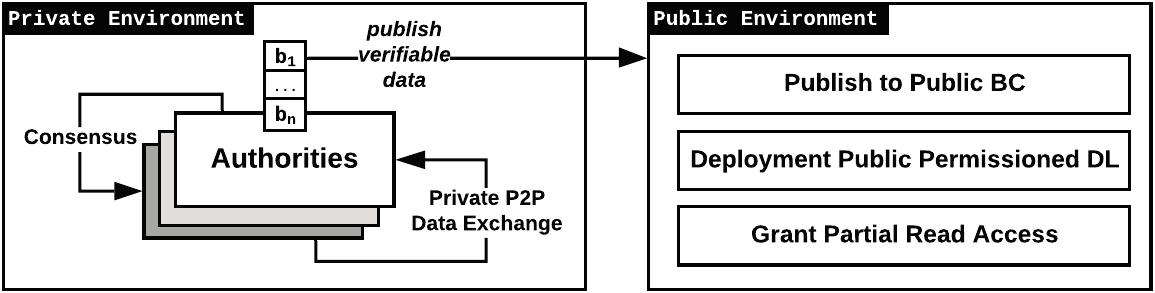}
    \caption{Proverum Architecture for \hpv}
    \label{fig:proverum}
    \vspaceshrinker
\end{figure}

Therefore, the Proverum architecture consists of a private and a public environment (\cf Figure~\ref{fig:proverum}). The private environment is formed by authorities collaborating as a private permissioned DL. Trusted authorities are authorized to participate in the DL; thus, they can be governmental authorities or private companies. Authorities execute a consensus protocol, validate transactions, group them into blocks while exchanging data in a Peer-to-Peer (P2P) fashion. Therefore, the private environment provides the distribution of the trust, including a privacy-preserving, verifiable audit log. The deployment of the public environment depends on requirements of public verifiability, which is achieved by either \1 publishing to a public BC, \2 deploying a public permissioned BC, or \3 granting partial read access to a subset of the private environment.

The public and the private environment require an information exchange interface. While the detailed interface specification depends on the architecture decisions taken, a fundamental requirement is that the private environment offers PKC capabilities; \ie each authority acts as a CA, thus can sign and thus authorize multiple keypairs for different target uses. Given that the private environment provides these capabilities (otherwise, the DL consensus and private P2P data exchange are practically impossible), a secure exchange interface can be implemented based on. Thus, cryptographic signatures and encryption schemes are then to be used for secure and authentic intra- and inter-authority data exchange.

Additionally, the interface can serve as a gatekeeper for private information, performing either security or integrity checks on the data or proofs to be published, hindering the leaks of private data. Depending on the use-case, the publishing process could include air-gapped processes which require offline signing.

\subsection{Hybrid Public Verifiability}
This work introduces a new definition of public verifiability: Hybrid Public Verifiability (\hpvnospace), which combines AV and PV, and thus implicitly includes the need for a private and public environment, \ie the requirement for a combination of private permissioned DLs and public permissioned BCs, clearly dividing the private audit trail (audit-trail for AV) from the verifiable public information (audit-trial for PV).

\begin{thm}{Hybrid Public Verifiability (\hpvnospace)}\\
A system offers \hpvnospace, when any individual can publicly verify the accuracy of all administrative procedures performed.
\end{thm}

Since PV as of Definition \ref{pv} includes the resistance against \textit{conspiracies of any size}, such an assumption is practically infeasible in many practical cases \eg Proverum applied to the currently deployed Swiss RPV, since many External Suppliers are trusted, as well as municipal authorities, employees and citizens counting paper ballots \cite{pvpf}. On the one hand, the underlying trust model of DLs and BCs depends on the consensus of these networks \ie on the architectural choices made. \textit{I.e.,} given a network operating with the Proof-of-Authority (PoA) consensus and $n$ authorities, a conspiracy of up to $(n/2)+1$ authorities can be tolerated. On the other hand, a Byzantine Fault Tolerance (BFT)-style consensus merely tolerates up to $1/3$ malicious nodes. Nevertheless, such an attack could easily be detected by other network participants (\ie other authorities). However, future approaches can surpass these limitations by using the Proverum architecture, achieving resistance to conspiracies of any size. Such a security property would require the deployment of additional cryptographic protocols embedded in the Swiss RPV process. 

\subsection{Private Environment Architecture}
The private environment of Proverum requires a high-level systems architecture, depicting different authorities and their access to various Smart Contracts (SC) deployed on DLs. Since identities of authorities maintaining the private environment are known, there is no need for a probabilistic consensus algorithm, \eg Proof-of-Work. Therefore, a suitable Byzantine Fault-Tolerant (BFT) consensus can be selected. Depending on the trust model, a single authority or a set of authorities is responsible for forming the consensus. 

\subsubsection{Network Architecture:} 
\begin{figure}[!ht]
    \centering
    \includegraphics[width=0.95\textwidth]{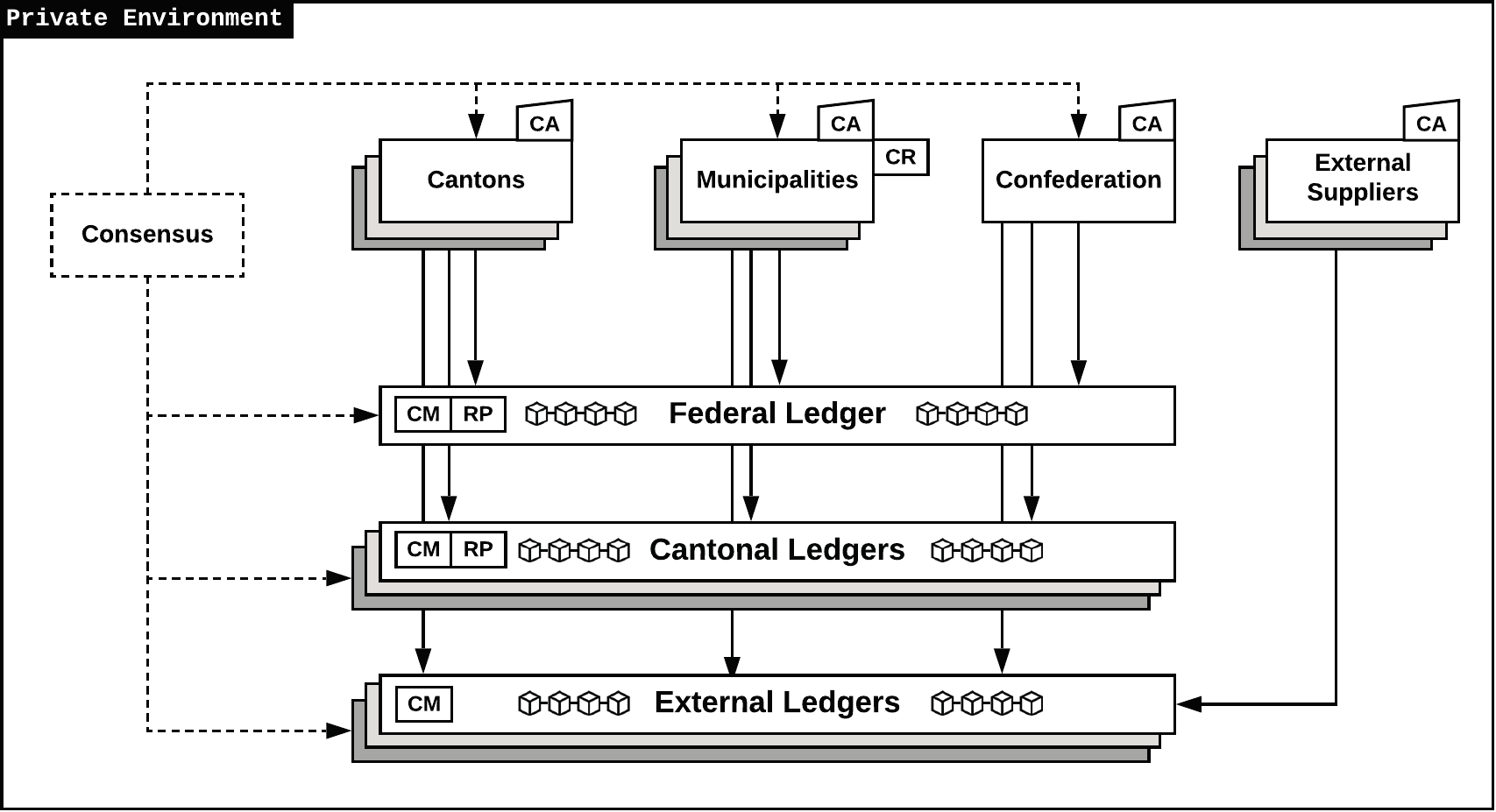}
    \caption{Proverum Private Environment Architecture}
    \label{fig:network-architecture}
    \vspaceshrinker
\end{figure}

Figure~\ref{fig:network-architecture} depicts the overview of the network architecture. Network participants are Cantons, municipalities, the Swiss confederation, and External Service Providers (ESP), \eg artifact manufacturers and the Swiss Post. For fault-tolerance reasons, each participant hosts at least two peers and acts as a Certificate Authority (CA) to issue certificates. 

Two purpose-specific SCs for \1 Citizen Management (CM) and \2 Result Publication (RP) allow keeping these SCs confidential among those involved. For instance, the ESP only has access to the \1 CM SC, while Swiss government authorities have access to the \2 RP SC. A consortium of all authorities forms the consensus. ESPs are not part of that consortium but have read access to the External DL. 

SCs and transactions are persisted on three different DLs. To one instance of a Federal Ledger, government entities are connected to. Cantonal Ledgers include one corresponding Canton and its subordinate municipalities, including the Swiss Confederation. In a country-wide deployment, 26 Cantonal Ledgers exist. Electoral authorities join several External Ledgers of ESPs on a municipal, Cantonal, and federal level. Since citizen data needs to be kept private at all times, each municipality operates a private data collection maintaining the Citizen Registry (CR) without exposing any details on a shared channel. 

\subsection{Public Environment Architecture}
\label{subsec:publicenv}
The public environment of Proverum can be supported by three different designs that are not mutually exclusive but can be combined to achieve public verifiability. For the public environment a public permissioned DL can be applied using a Proof-of-Authority (PoA) consensus mechanism since PoA reduces the proof efforts, meaning that only participants that have been given authorization can produce blocks \cite{caiv}. As well, different trust models apply to each design. Thus, depending on the specific requirements to be met, the Proverum architecture allows for all three options to be followed. 
Each design option (summarized as of Table~\ref{tab:publicenv}) achieves public verifiability, whereas \1 the publication to a public BC mainly relies on the data to be authentic. For \2, the deployment overhead is large, but the environment achieves the best public verifiability, showing a fully transparent audit trail that the public can audit and authorities can control. The option \3 opens up the Application Programming Interface (API), is easily implemented, but also achieves the weakest notion of public verifiability, since a single trusted endpoint represents a single point of failure. 

\begin{table}[!ht]
\newcommand{\centered}[1]{\begin{tabular}{@{}l@{}} #1 \end{tabular}}
\newcommand{\STAB}[1]{\begin{tabular}{@{}c@{}}#1\end{tabular}}
\caption{Comparison of Proverum's Public Environments}
\label{tab:publicenv}
\begin{tabular}{|P{0.15\textwidth}|P{0.2\textwidth}|P{0.2\textwidth}|P{0.2\textwidth}|P{0.2\textwidth}|}
\cline{2-5}
\multicolumn{1}{c|}{} & \multicolumn{4}{c|}{\textsc{Property}} \\ 
\cline{2-5}
\multicolumn{1}{c|}{} & \textbf{Trust Model} & \textbf{Key Management} & \textbf{Deployment Effort} & \textbf{Public Verifiability}\\
\hline
\textbf{1. Publish to Public BC} & Trust in \1 permissionless consensus, \2 developers of BC, \3 authorities & Keypair per public BC and per authority in private environment & Medium & To query SC \\
\hline
\textbf{2. Public Permissioned DL} &  Trust in authorities & Keypair per authority in public and private environment & High & To query full BC \\
\hline
\textbf{3. Grant Partial Read Access} & Trust in Authorities & Keypair per authority for private environment &  Low & To query API \\
\hline
\end{tabular}
\vspaceshrinker
\end{table}

\begin{enumerate}[align=left]
\setlength\itemsep{0.2em}
    \item \textbf{Publishing to a Public BC} enables the publication of data and proofs (\eg cryptographic hashes of ERs, aggregated voting results, or Non-Interactive Zero-Knowledge Proofs) by directly signing cryptographically a transaction containing these data, and broadcasting them to the respective public BC. This requires appropriate key management on the side of publishing authorities since every public BC requires the management of public/private keypairs to perform the transaction signature.
    Alternatively, one can rely on~\cite{bifrost} to send and retrieve data within a transaction to multiple BCs without being locked into a single platform. With such a BC-agnostic platform, the immutability of public BCs is exploited with less complex barriers. The publication on a public permissionless BC, such as Bitcoin or Ethereum, offers transparency and public verifiability. 
    The trust model depends on the specific public BC chosen since they vary in their selection of consensus algorithms. However, since public permissionless BCs do not rely on any Trusted Parties, the public can be assured that data retrieved was stored in a censorship-resistant and tamper-proof manner~\cite{bcInTheWings}. 
    \item \textbf{The Deployment of a Public Permissioned DL} requires additional operations. This is similar to the approach used in \cite{caiv}, where a Public Permissioned DL serves as a publicly readable PBB. Therefore, the trust model is comparable to the private environment, since the DL is formed by trusted authorities. The main difference compared with the publication on a public BC is that authorities can censor transactions in a permissioned setting and change the content of the DL. However, since the public can read all data persisted on that DL, such changes could be observed and audited by the public, which is allowed to participate as a full node, persisting and duplicating the DL. From an economic perspective, the deployment of a public permissioned DL requires an operational effort to maintain and operate. 
    \item \textbf{Granting Partial Read Access} to a subset of the private environment can be offered by providing a public interface, which can be queried by anyone. However, providing a centralized point of failure to a distributed system providing verifiability is not optimal, since it separates private transactions from the public, \eg by only giving access to a public interface that can be queried. The trust model here resembles a central server, since no distribution of trust is taking place towards the public. Economically, this option is most
\end{enumerate}
 
\section{Prototypical Implementation of Proverum}
\label{sec:implementation}
For the instance of the private environment being prototyped HLF is applied. Therefore, SCs are referred to as \textit{Chaincodes} and DLs among a subset of authorities are referred to as \textit{Channels} \cite{hlfdocs}.
The current prototype focuses on the implementation of the private environment. The source code is available in \cite{proverum}. 

\begin{figure}[!t]
    \centering
    \includegraphics[width=0.95\textwidth]{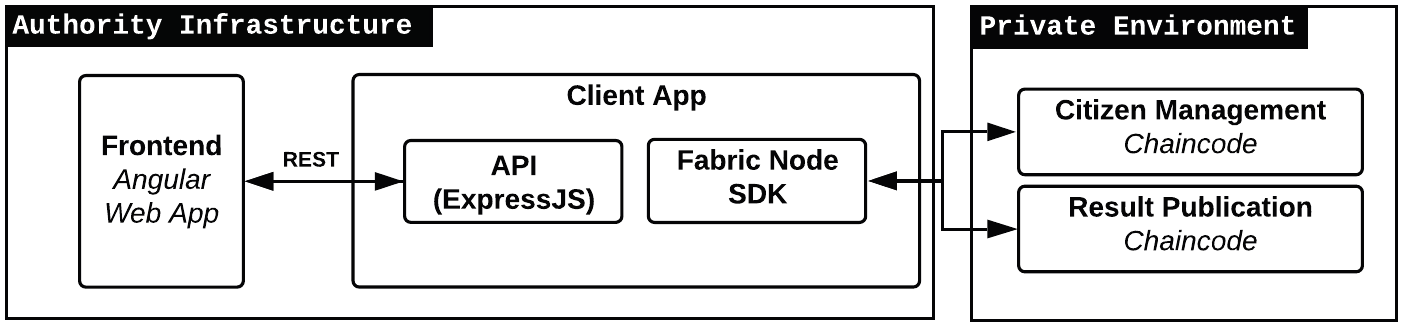}
    \caption{Overview of the Proverum Prototype}
    \label{fig:prototype-layers}
    \vspaceshrinker
\end{figure}

The prototype shows two components executed on the infrastructure of the authority: \1 a frontend Web application implemented in Typescript, using Angular in a Model-View-Controller architecture~\cite{angular}. The frontend serves a dashboard showing the current network status. It contains input forms and pop-up dialogs to enable authorities to submit transactions easily and trigger chaincode calls through the Client App (\cf Figure~\ref{fig:prototype-layers}).
Second, \2 the Client App interacts through the Fabric Node SDK with the BC infrastructure and encapsulates the various chaincode calls within REST endpoints based on the ExpressJS Web application framework.

In HLF, the global state of the DL is separated based on the chaincode that accesses it. Therefore, two world states are maintained, one for the CM and one for the RP chaincode. The separation of states is required, since both chaincodes are distinct and involve different stakeholders. All chaincodes are written in Javascript for the node.js runtime environment, using the HLF Chaincode Shim API to process DL states and communicate with other peers~\cite{hlfdocs}. The CM chaincode contains Create, Update, Update, and Delete functionality to manage CRs and the generation and sharing of the ERs. Additionally, the CM chaincode allows for the generation of Non-Interactive Zero-Knowledge Proofs, proving the eligibility of voters contained in the ER. 

\subsubsection{Prototype Network Setup:}
\begin{figure}[!ht]
    \centering
    \includegraphics[width=1\textwidth]{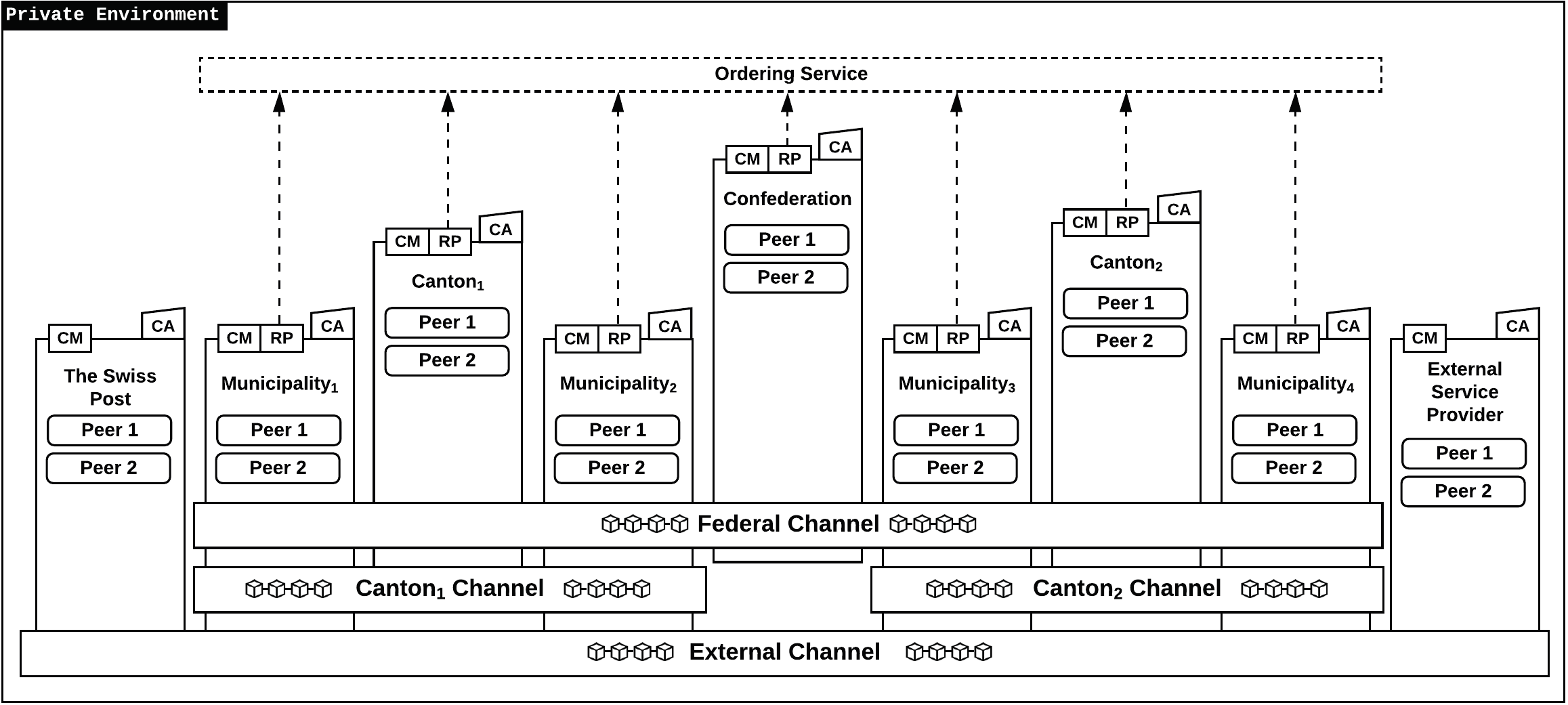}
    \caption{Prototype Network Architecture}
    \label{fig:prototype-network}
\end{figure}

The prototype network is composed of the confederation, two Cantons, and two municipalities per Canton (\cf Figure~\ref{fig:prototype-network}). Additionally, the Swiss Post and a single ESP (\eg a voting artifact supplier) serve as a provider for all municipalities. A total of 2,202 municipalities is served in a country-wide deployment, split up among 26 Cantons, including different ESPs. Each stakeholder hosts two peers for fault tolerance purposes. HLF forms consensus by using an Orderer, a dedicated node arranging transactions in a deterministic fashion \cite{hlfdocs}. All authorities of the consortium are part of the Ordering Service (\cf Figure~\ref{fig:prototype-network}). Hence they can serve as Orderer and create Channels.

The CM chaincode is required on all peers, while the RP chaincode is only maintained on government peers to keep it separated from any external channels. Thus, four different channels exist in the network. On the federal channel, where government authorities are connected, Cantonal channels are occupied by the Canton and its two subordinate municipalities. The municipalities 1 to 4 and ESPs maintain private citizen data in an instance of a database. 

\subsubsection{Data Model:} 
\begin{figure}[!ht]
    \centering
    \includegraphics[width=0.95\textwidth]{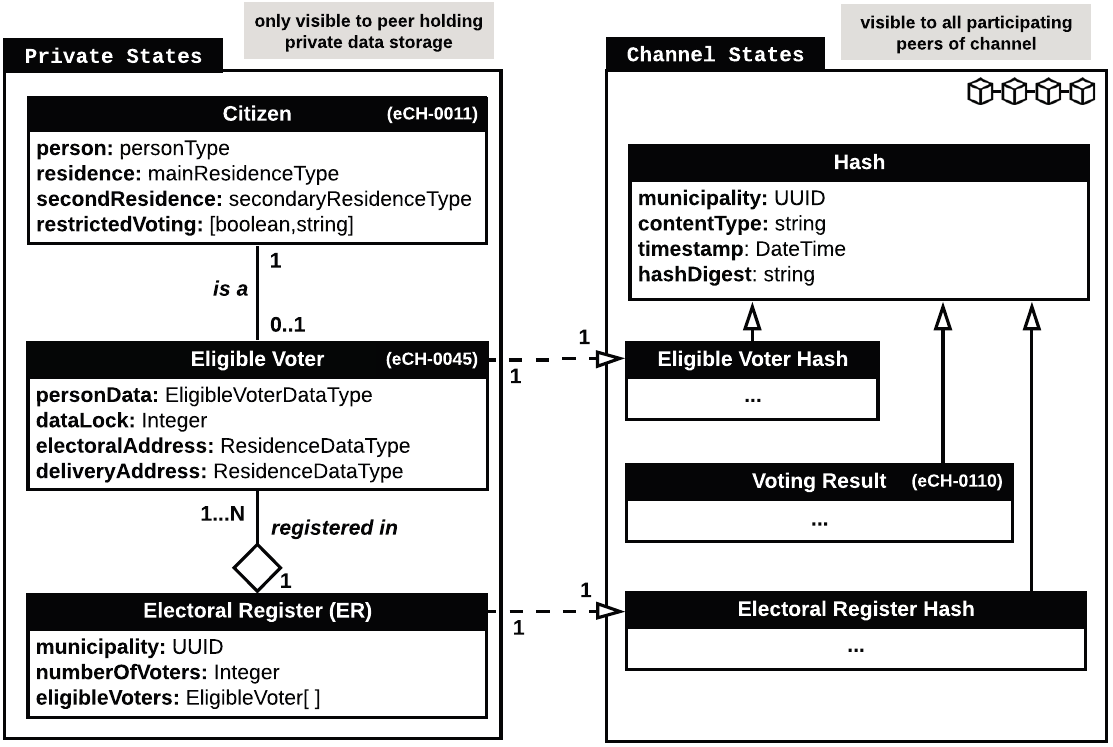}
    \caption{Class Diagram}
    \label{fig:classdiagram}
\end{figure}

It is derived from the eCH eGovernment \cite{ech}. Information is encapsulated in a state object, which can be queried with a state key. Citizen states are modelled according to \texttt{eCH-0011}. These fields contain basic personal information encapsulated in data types defined by, \texttt{eCH-0044}, \texttt{eCH-0045}, and are depicted in Figure~\ref{fig:classdiagram}. The citizen state contains a method to invoke a voting restriction in case a person becomes non compos mentis. Citizen states may have a corresponding voting citizen state counterpart, if the citizen is an eligible voter, implying Swiss citizenship, adulthood, a main residency in the registered municipality, and no voting restrictions. Therefore, eligible voters show an aggregation relationship to the voting list state representing the ER. These states are all private; thus, they are off-chain, persisted in the private data storage of the municipality. Only the corresponding cryptographic hash digest of individual eligible voters, the ER, and plain voting results are persisted on-chain.

\section{Use Cases}
\label{sec:usecases}
Three Use Cases (UC) are defined to evaluate the prototype of Proverum. Each UC shows different operations and actions being performed by those stakeholders involved. \textit{UC1} focuses on the IdM of citizens, in detail, the relocation of citizens. \textit{UC2} and \textit{UC3} focus on the support of voting and electoral processes in the Swiss RPV, especially the generation and validation of ERs and the voting and election result publication process, respectively.

\subsubsection{UC1 Citizen Relocation:}
\label{subsec:uc1}
\begin{figure}[t!]
    \centering
    \includegraphics[width=0.95\textwidth]{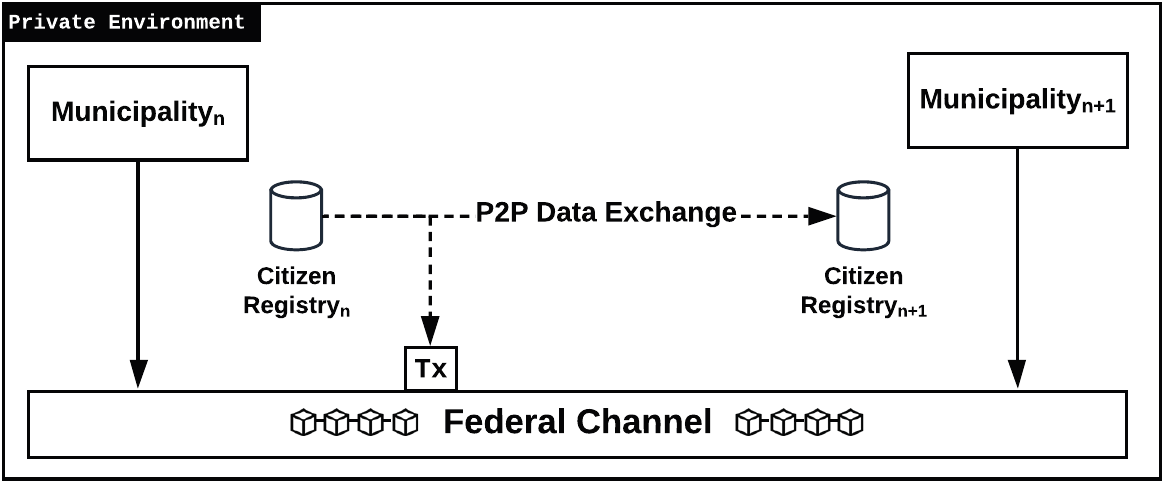}
    \caption{\textit{UC1:} Citizen Relocation}
    \label{fig:uc1}
    \vspaceshrinker
\end{figure}

In the event of a citizen relocating from municipality $n$ to municipality $n+1$, citizen data needs to be \1 sent first from $n$ to $n+1$. Then \2 the original municipality $n$ deletes all records. Since citizen data need to remain private to these municipalities, they are shared P2P via a gossip protocol (\cf Figure~\ref{fig:uc1}). The Ordering Service is not involved and cannot access private data; only peers of authorized municipalities have access. Thus, private citizen data is not included in the transaction to the orderer \cite{hlfdocs}. Only a cryptographic hash digest of the private data is endorsed, ordered, and persisted on DLs of peers on the Federal Channel. Finally, the hash serves as evidence of the transaction and is used for state validation and can be used for audit purposes, too.

\subsubsection{UC2 Generation and Validation of Electoral Registers:}
\label{subsec:uc2}
\begin{figure}[t!]
    \centering
    \includegraphics[width=0.95\textwidth]{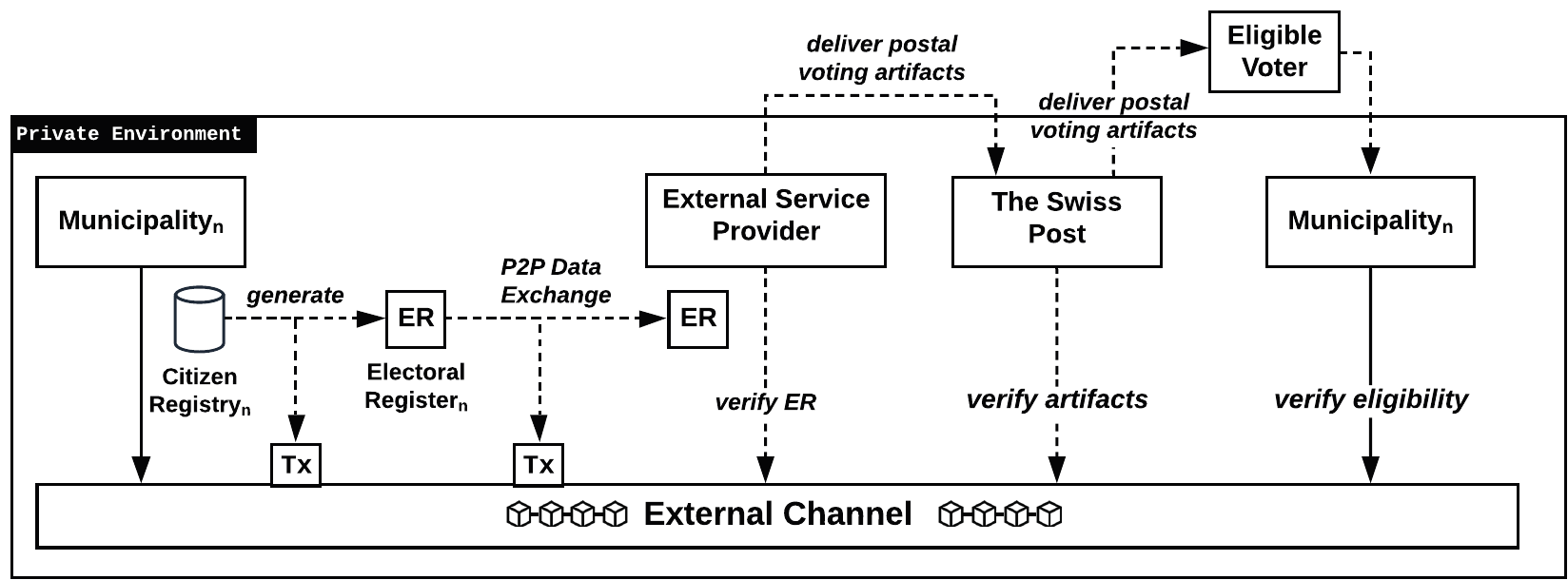}
    \caption{\textit{UC2:} Generation and Validation of Electoral Registers (ER)}
    \label{fig:uc2}
    \vspaceshrinker
\end{figure}

As depicted within Figure \ref{fig:uc2}, when a new vote is scheduled, municipalities generate the respective ER and the subset of the CR containing all eligible voters. The CM chaincode manages the generation of the ER. The CR is checked and excludes all citizens that \1 do not possess Swiss citizenship, \2 are not older than eighteen, or \3 do not have their primary residence within the given municipality. 

The ER is exchanged to either one or multiple ESPs, which will prepare and print physical voting artifacts (\eg voting envelope, paper ballots, and paper ballot envelope). Additionally, the municipality publishes cryptographic hashes of the complete ER and hashes of each voter on the external channel, serving as a time-stamped proof for verification or audits. When the manufacturer receives these private data, they will be checked against hashes published to ensure that the data transmitted is correct. 
This allows the Swiss Post (SP) to verify artifacts upon reception. In detail, as soon as the SP receives these artifacts, each envelope is scanned to verify, whether the hash of each ballot is contained in the ER, also verifying the completeness of the batch by computing the hash of the complete list. This allows for the detection of loss, theft, or incomplete production of voting artifacts. Voters can gain public verifiability by seeing a read operation on the External Channel through personalized hyperlinks, \eg being embedded on the postal voting artifacts as a QR code (Quick Response).
\subsubsection{UC3 Accumulation and Publication of Preliminary Results:}
\label{subsec:uc3}
Authorities often use insecure communication channels (\eg telephone, fax, or non-secured electronic mail) to exchange preliminary results within Switzerland. From a local and global perspective, election laws do not require any guarantee regarding the integrity of electronically transmitted results, which bears the potential for attacks \cite{sommer2019cyberrisks}. Thus, UC3 defines a secured process to assure integrity and auditability for preliminary as well as final election results.

\begin{figure}[!t]
    \centering
    \includegraphics[width=0.95\textwidth]{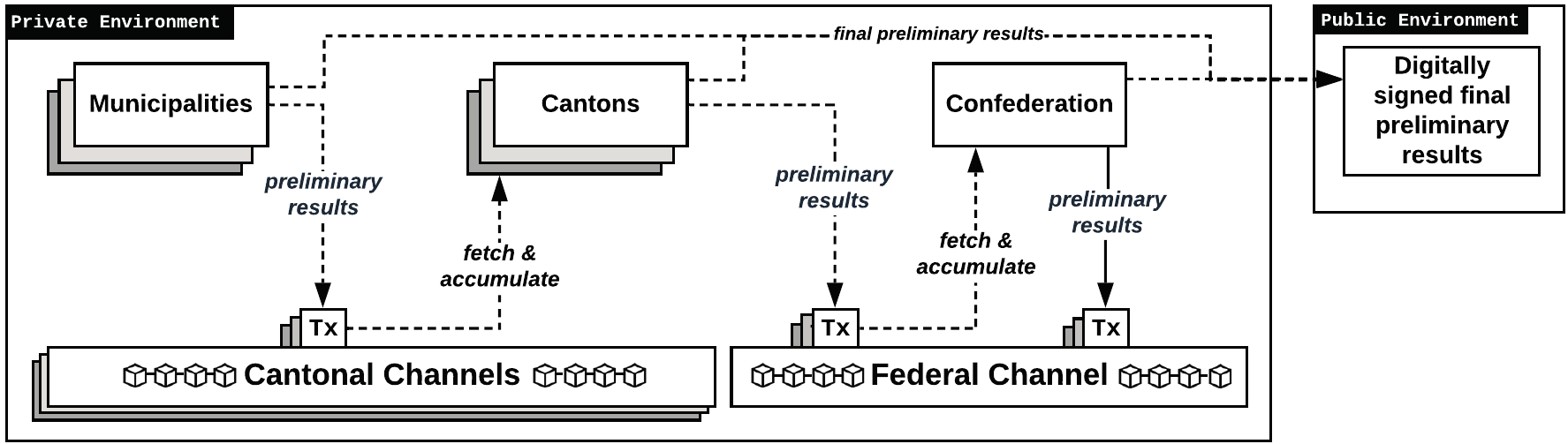}
    \caption{\textit{UC3:} Secured Accumulation and Publication of Preliminary Results}
    \label{fig:uc3}
    \vspaceshrinker
\end{figure}

Therefore, UC3 details the propagation of preliminary results from a municipality to the Cantonal level by publishing results through Cantonal Channels (\cf Figure \ref{fig:uc3}). In turn, Cantons perform plausibility checks, and, if necessary, trigger manual recounts. Furthermore, Cantons publish accumulated results from all municipalities and publish these on the Federal Channel, which the federal government will assemble to tally the preliminary results.

Municipalities, Cantons, and the confederation digitally sign final preliminary results and publish the data in the public environment, especially to a public permissioned BC or a public BC, which enables public verifiability. While still relying on the integrity of these individual results published by the different authorities involved, UC3 reaches the verifiability for the preliminary tally publication. Instead of relying on insecure communication channels, the preliminary results' authenticity and integrity are cryptographically secured.
\section{Evaluations}
\label{sec:evaluation}
The evaluation of Proverum is grouped according to Threat Events (TE) of the Swiss Postal Voting Process Flow (PVPF) as outlined in \cite{pvpf}. Table \ref{tab:RAA} lists all TEs, alongside the respective Mitigation (M) techniques, which are considered relevant here. Fundamental elements contributing toward mitigation include: \1 the Public-Key Infrastructure paired with \2 the use of the private Proverum environment as a distributed, immutable audit trail on a private permissioned DL, \3 the public environment for public verifiability, and \4 the use of SCs to automate verifiable and distributed workflows for a decentralized IdM. 

\begin{table}
\centering
\caption{Threat Events and Mitigation Techniques of the Swiss RPV \cite{pvpf}}
    \begin{tabular}{|c|l|l|}
    \hline
    \small
    \textbf{TE} & \textbf{Description}  & \textbf{Mitigation} \\
    \hline 
    \textbf{TE1} & Delay production of physical artifacts & \textit{no Mitigation}  \\
    \hline
    \textbf{TE2} & ER master records  & \textit{Immutable Audit Trail}  \\
    \hline
    \textbf{TE3} & ER data snapshot & \textit{Cryptographically signed Tx}  \\
    \hline
    \textbf{TE4} & Forge physical artifacts  & \textit{Eligibility Verifiability Mechanism} \\
    \hline 
    \textbf{TE5} & Steal assembled VEs before dispatch & \textit{Artifact Verification Mechanism}  \\
    \hline 
    \textbf{TE6} & Re-route VEs & \textit{Immutable Audit Trail} \\
    \hline
    \textbf{TE7} & Steal VEs from voter letterboxes & \textit{Report, Blacklist and Re-Issuance}  \\
    \hline
    \textbf{TE8} & Steal VEs from municipal letterbox & \textit{Report, Blacklist and Re-Issuance}  \\
    \hline
    \textbf{TE9} & Re-route VEs & \textit{Immutable Audit Trail}  \\
    \hline
    \textbf{TE10} & Cast stolen or forged VEs & \textit{Artifact Verifiability Mechanism}  \\
    \hline
    \textbf{TE11} & Access stored VEs & \textit{Immutable Audit Trail}  \\
    \hline
    \textbf{TE12} & Manipulate tallying & \textit{Smart Contract \& Audit Trail}  \\
    \hline
    \textbf{TE13} & Manipulate final tally & \textit{Smart Contract \& Audit Trail} \\
    \hline
    \textbf{TE14} & Initiate premature destruction & \textit{Cryptographically signed Tx} \\
    \hline
    \end{tabular}
    \label{tab:RAA}
    \vspaceshrinker
\end{table}

\begin{description}[align=left]
\setlength\itemsep{0.2em}

\item[TE1] refers to targeted cyberattacks on ESPs or municipal information systems (\eg in the form of Distributed Denial-of-Service attacks). There does not exist a single explicit mitigation strategy, since cybersecurity best-practices need to be applied across all processes and the entire infrastructure involved.

\item[TE2] refers to tampering with CR master records. As of today, municipalities and Cantons manage their CR in centralized information systems. In contrast, Proverum records every change in a CR master record in private DL's immutably records, which simplifies the audit, detects errors, and an indication of fraudulent changes in the CR. Subsequently, in the ER.

\item[TE3] describes the risk of tampering with ER snapshot data in transit from the municipality to an ESP, \eg to print VSCs, and assemble VEs. As outlined for UC2, the ER is now persisted on the Cantonal Ledger in a digitally signed transaction, which is persisted on the immutable, tamper-proof DL. 

\item[TE4] describes the forgery of physical voting artifacts, which requires specific knowledge of the Swiss PVPF and access to various digital templates. With Proverum the eligibility verifiability is assured, since each ballot cast is cross-checked with the ER, which was published on the Cantonal Ledger (\cf UC2, Figure~\ref{fig:uc2}). To consider an attack successful, the adversary requires access to the authorities to create fake identities, which would need to be included into the CR and ER, in order to pass verification checks during tallying.

\item[TE5] refers to the theft of assembled VEs before they are dispatched. As proposed within UC2, the SP verifies the receival of all VEs, which assures that not a single VE is missing. In case of theft, the SP will realize misses and reports the theft to Proverum, which automatically \1 triggers a re-issuance of a new ER with newly salted hashes, so the hashes are different from the ones issued previously and \2 all the stolen VEs are invalidated in an explicit blacklist transaction. These blacklists are checked during the tallying phase. 

\item[TE6] describes the re-routing of VEs, which assumes an adversary to have access to internal systems of the SP and may require co-conspiring postal employees. Proverum offers the immutable audit trail. Hence, the detection of such attacks is eased, since irregularities in the delivery of VEs can be detected by any member of the respective DL (\eg the Cantonal Ledger).

\item[TE7] describes the theft of yet unused VEs from voters' letterboxes. Even if all prior steps of the PVPF were executed correctly, the mitigation of this attack depends on the voter to realize a theft and \1 report it to the municipality's election office, which can \2 re-issue a second set of voting artifacts to be used to cast a valid, eligible vote. The stolen VEs are \3 invalidated in an explicit blacklist transaction and newly created artifacts are registered on the Cantonal Ledger.

\item[TE8] describes the theft of VEs from the municipal letterbox, which can contain \1 VEs returned by SP or \2 manually returned by voters. This distinction is essential, since manually returned VEs are not reported as \textit{returned} in the shared DL. Therefore, if a voter knows he/she returned the VE and wants to verify the audit trail and the VE did not get registered (and reported as \textit{received}) by the election office. Similarly to TE7, this needs to be \1 reported, \2 old artifacts are blacklisted and \3 artifacts are re-issued. Proverum records all actions in transactions on the DL, thus creating an audit trail. 

\item[TE9] refers to the re-routing of VEs as soon as they cast and return the VE via SP. This is handled as for TE6. 

\item[TE10] refers to the casting of stolen or forged VEs. Similarly to TE4, an adversary is required to refer to the casting of stolen or forged VEs and an adversary would require access to the CR.

\item[TE11] describes the threat of storage access to VEs cast within the respective municipality and, thus, is not a threat that can be mitigated by purely digital means since it requires trust in authorities. In contrast, Proverum addresses the risk of theft of VEs by logging and publishing all incoming VEs on the DL, thus, making a disappearance of any VE detectable. 

\item[TE12] refers to the risk of manipulation during tallying. Although the manual counting mechanisms are out of scope of Proverum, the private environment could be used to record either \1 individually counted tallies or, as implemented in UC3, preliminary results. The SCs enforce the correct counting of the results entered and allow all peers in the DL to verify the correct execution, thus, distributing the trust among all authorities. Further, instead of using insecure communication channels, cryptographically signed and broadcast transactions are persisted on the DL.   

\item[TE13] describes the manipulation of the final tally. Since most Cantons use proprietary software to handle vote transmission for preliminary results \cite{sommer2019cyberrisks}, an adversary can tamper with these results. Similarly to TE12, the cryptographically signed transaction assures that results can not be tampered with, since they are included in the immutable and distributed DL.

\item[TE14] describes the triggering of premature destruction. Municipalities await a formal message over insecure communication channels to receive the approval to destroy all physical voting artifacts. For instance, a potential attack could send e-mails to various municipalities, triggering premature destruction. Proverum mitigates this by using the DL to communicate only authenticated, and verifiable destruction commands to municipalities. 

\end{description}

Thus, the Proverum architecture is generally an effective combination of a private and public environments to achieve \hpv and serves as a highly suitable approach for a decentralized Identity Management (IdM) of citizens, including unforgeable states. The private environment is based on a private permissioned DL, which forms the immutable audit-trail, enables the decentralized IdM, and allows for an automated process with SCs of otherwise manual and error-prone analog processes. The public environment achieves transparency by using public (permissioned DLs) to reach public verifiability. 

The artifacts are verified again as soon as the votes were cast and returned to the municipality. Forged voting artifacts can be easily detected since the cryptographic hash digest generated in the first step was published on the channel and verified. However, it is still possible for an adversary to steal artifacts (and then cast them) when they are in transit between the Swiss Post and the Eligible Voter. Even in the case of theft, it would be possible to detect when many voters suddenly ask for an additional set of voting material.

\section{Summary and Conclusions}
\label{sec:conclusion}
Since verifiability and trust are fundamental for transparency in voting processes, the explicit application of the Swiss Remote Postal Voting (RPV) case serves for Proverum as a real-world example with various trusted stakeholders interacting via insecure communication channels \cite{pvpf}. The exploitation of prior work (analysis and a risk assessment leading to Threat Events, TE) lead to the new design of Proverum and its prototypical implementation, which laid via three Use Cases (UC) the technical basis to achieve public verifiability in the Swiss RPV approach.

Like PBBs in electronic voting, the Proverum architecture provides an immutable audit trail with multiple permissioned DLs for a clear distinction of a private environment and public BCs. The prototypical implementation highlights the practical feasibility of using a Web-based frontend application, which can be easily operated by the authorities, and which interacts with a Client App connecting to the permissioned DL. Note that the publication of verifiable information (\eg cryptographic hashes of Electoral Registers (ER) or Non-Interactive Zero-Knowledge Proofs, proving the eligibility of voters listed) is crucial to building trust in any digital process. Thus, trust is not only distributed and enforced by the DL, but Proverum serves as a comprehensive platform to enable transparency as well as administrative and public verifiability. Likewise, the integrity of any final tally is only trusted, when public verifiability mechanisms are provided. 

The application of the Proverum architecture in the Swiss RPV cases has proven to serve as an effective and technically feasible approach in a governmental setting to provide \1 trust, \2 integrity, \3 transparency, \4 \hpvnospace, and \textit{(v)} an architecture for a decentralized IdM. In that sense, the exploitation of public permissionless Blockchains (BC), as well as public permissioned Distributed Ledgers (DL), makes Proverum ready for other electoral processes as well or a sharing of information in the health-care sector as stated. 

\section{Future Work}
Future work has already started and focuses on \1 the proposal of novel cryptographic protocols to be used in the Swiss RPV context, leveraging the Proverum approach with immutable evidence storage. Further steps cover \2 the formal definition of the protocol used for enabling \hpv in the Swiss RPV system, and \3 additional prototypical implementation steps and performance evaluations concerning the scalability of those use-cases implemented.

\section*{Acknowledgements}
This paper was supported partially by \textit{(a)} the University of Z\"urich UZH, Switzerland and \textit{(b)} the European Union's Horizon 2020 Research and Innovation Program under Grant Agreement No. 830927, the CONCORDIA Project. %

\bibliographystyle{splncs03}
\small 
\bibliography{refs}

\ \\
All links were last followed on August 20, 2020.
\end{document}